\documentclass[12pt]{article}
\usepackage{graphicx}
\newcommand{\mapright}[2]
{\mathop{\hbox to 8mm{\rightarrowfill}}
\limits^{\scriptstyle #1}_{\scriptstyle #2}}

\begin{document} %
\begin{flushright}
{KOBE-TH-04-07}\\
\end{flushright}
\renewcommand{\thefootnote}{\fnsymbol{footnote}}
\vspace*{3mm}
\begin{center}
{\large \bf
Gauge Symmetry Breaking with a Large Mass Hierarchy\footnote[2]{Talk at Summer Institute 2004 (SI 2004), Fuji-Yoshida, Japan (12-19 August, 2004). This work was supported 
in part by the Grant-in-Aid for Scientific Research (No.15540277) 
by the Japanese Ministry of Education, Science, Sports and Culture. }} 
\vspace{8mm} \\
\renewcommand{\thefootnote}{\arabic{footnote}}
Tomoaki Nagasawa
${}^1$ 
and 
Makoto Sakamoto
${}^2$
\vspace*{4mm} \\
{\small \it
${}^1 $Graduate School of Science and Technology, Kobe University,
Rokkodai, Nada, \\ Kobe 657-8501, Japan
\\
${}^2 $Department of Physics, Kobe University,
Rokkodai, Nada, Kobe 657-8501, Japan
}
\vspace{6mm}
\\
\begin{minipage}[t]{130mm}
\baselineskip 4mm 
{\small
We propose a higher dimensional scenario to solve the gauge hierarchy problem. 
In our formulation, a crucial observation is that a supersymmetric structure 
is hidden in the 4d spectrum of any gauge invariant theories 
with compact extra dimensions and that the gauge symmetry breaking is 
directly related to the supersymmetry breaking.
}
\end{minipage}
\end{center}
\vspace{5mm}
\noindent
%
\baselineskip 5mm 
 
Recently, new possibilities of Grand Unified Theories (GUTs) 
in the context of higher dimensional field theories have 
extensively been explored to avoid common problems of four-dimensional GUTs.
The GUT symmetry breaking scale is naturally given 
by the inverse length of the extra dimensions, 
but the origin of the weak scale is still a mystery.

In this paper, we propose a higher dimensional scenario 
to solve the hierarchy problem and explain how to get 
4d masses much smaller than any scales of fundamental 
higher dimensional gauge theories without a fine tuning.
Our mechanism to get hierarchically small masses is based on 
the idea that non-perturbative supersymmetry breaking can produce 
an exponentially small (supersymmetry breaking) mass scale such as
\begin{equation}
	M \times e^{-\kappa}  <\hspace{-,4em}<\hspace{-,4em}<  M,
\end{equation}
where $M$ denotes a typical mass scale of the system and 
the exponential factor $e^{-\kappa}$ comes from a non-perturbative effect 
of supersymmetry breaking.
Fortunately, a supersymmetric structure is already built in {\it any} 
gauge theories with compact extra dimensions \cite{1,2}, 
and we can use it in our formulation.
Then, the supersymmetry breaking immediately implies gauge symmetry breaking 
with the hierarchically small mass scale given by (1), 
which is a desired situation in constructing GUT models 
without the gauge hierarchy problem.

Let us begin to give a plausible argument for the statement 
that a supersymmetric structure is actually hidden 
in any gauge invariant theories with compact extra dimensions.
To this end, let us consider a 4+1-dimensional gauge field 
$A_M(x,y)=(A_{\mu}(x,y),$ $A_5(x,y))$.
Expanding them into KK modes $(A_{\mu , n } (x), A_{5 , n} (x) )$, 
we find that the answers to the questions 
\begin{itemize}
	\item why there exists a 4d massless vector $A_{\mu , 0}$\,? 
\vspace{-3mm}	
	\item why massive modes $A_{5 , n} $ are unphysical\,? 
\vspace{-3mm}	
	\item why a massless mode $A_{ 5, 0 }$ (if exists) is physical\,?
\end{itemize}
can equally be explained in the languages of gauge symmetry 
and supersymmetry as follows:
\begin{table}[h]
      \begin{center}
       \begin{tabular}{|c|c|}
         	\hline
			gauge symmetry point of view 
			&
			supersymmetry point of view
			\\
		\hline 
		\hline
			The 4d gauge invariance ensures that
			&
			Supersymmetry ensures that
			\\
			the zero mode $A_{\mu , 0 }$ is massless.
			&
			the vacuum state has zero energy.
			\\
			\hline
			All massive modes $A_{5,n}$ can be absorbed 
			&
			The fact that the longitudinal modes of 
			\\
			into the longitudinal modes of $A_{\mu ,n}$  		
			&
			$A_{\mu ,n}$ form superpartners with $A_{5,n}$ ensures 
			\\
			by gauge transformations.
			&
			that $A_{5,n}$ are gauged-away.
			\\
			\hline
			Gauge transformations cannot remove  
			&
			Zero modes are special in SUSY 
			\\
			the zero mode $A_{5,0}$ because 
			&
			because they do not, in general,
			\\
			$A_5 \to A_5 +\partial_y \Lambda$.
			&
			 form supermultiplets.
			 \\
			\hline
	          \end{tabular}
     \end{center}
\end{table}
\\
The above observations strongly suggest that the gauge symmetry is 
closely related to the supersymmetry.
In fact, we can prove that an $N=2$ supersymmetric quantum-mechanical 
structure exists in any gauge invariant theories 
with compact extra dimensions \cite{2}.
An explicit exercise will be given below (see also Ref. \cite{1}).

In order to demonstrate how a hierarchically small mass arises 
in a higher dimensional gauge theory, let us consider 
a 4+1-dimensional gauge invariant theory on an interval 
$( 0 \le y \le L )$ with the action
\begin{equation}
	S=\int d^4 x \int_0^L dy \Delta(y) \left\{
		-\frac{1}{4} F_{MN}^a (x,y) F^{aMN}(x,y) \right\},
\end{equation}
where $\Delta(y)$ is some weight function depending only 
on the 5th coordinate $y$.
We here take $\Delta(y)$ to be of the form 
$\Delta(y)=\exp \left( -\frac{2\kappa}{L}y\right)$, 
as an illustrative example.

To obtain the 4d effective theory, we expand the gauge field $A_{\mu} (x,y) $ 
as $A_{\mu}(x,y) =\displaystyle{\sum_n} A_{\mu , n} (x) f_n(y)$, where
\begin{equation}
	-\left[
		\frac{1}{\Delta(y)} \partial_y \Delta(y) \partial_y \right]
			f_n(y) =m_n^2 f_n(y),
\end{equation}
while we expand 
$A_5(x,y)$ as $A_5(x,y)=\displaystyle{\sum_n} A_{5,n}(x) g_n(y)$, where
\begin{equation}
	-\left[
		\partial_y \frac{1}{\Delta(y)} \partial_y \Delta(y) \right]
			g_n(y) =m_n^2 g_n(y).
\end{equation}
Here, $m_n$ corresponds to the mass of the 4d gauge boson $A_{\mu, n}$.
It is easy to see that the eigenfunctions $f_n(y)$ and $g_n(y)$ 
are mutually related as
\begin{equation}
	Q \left(
		\begin{array}{c}
			f_n \\
			g_n 
		\end{array}
	\right)
	=m_n 
	\left(
		\begin{array}{c}
			f_n \\
			g_n 
		\end{array}
	\right), 
	\qquad 
	Q=\left(
		\begin{array}{cc}
			0 & -\frac{1}{\Delta(y)} \partial_y \Delta(y) \\
			\partial_y &0 
		\end{array}
	\right).
\end{equation}
The differential operator $Q$ can be identified with a supercharge 
and in fact the above two differential equations, (3) and (4), 
can simply be combined into the form 
\begin{equation}
	H \left(
		\begin{array}{c}
			f_n \\
			g_n 
		\end{array}
	\right)
	=m_n^2 
	\left(
		\begin{array}{c}
			f_n \\
			g_n 
		\end{array}
	\right),
	\qquad 
	H=Q^2,
\end{equation}
which is known as an $N=2$ supersymmetric quantum mechanics \cite{3}.

This is not, however, the end of the story.
Since the extra dimension has two boundaries at $y=0$ and $L$, 
we have to specify some boundary conditions (BC's) there.
The BC's should be compatible with gauge invariance.
This requirement may be replaced by the statement that 
the BC's should be compatible with supersymmetry: 
for any wavefunction $\Psi (y)$ that obeys the required BC's, 
the state $Q \Psi (y)$ has to obey the same BC's, 
otherwise the supercharge $Q$ is ill defined.
It turns out that the above requirement severely restricts 
allowed BC's, and in fact only two types of BC's at each boundary 
are compatible with the supercharge $Q$ \cite{4}:
\begin{eqnarray}
	&&{\rm Type\ N} : \partial_y f_n(y)=0 \quad {\rm and} \quad g_n(y)=0, \\
	&&{\rm Type\ D} :  f_n(y)=0 \quad {\rm and} \quad \partial_y \left( \Delta(y) g_n(y)\right)=0, 
\end{eqnarray}
at $y=0, L$. 
It follows that there are 4 possible models with type (N,N), (D,D), (D,N) and (N,D), 
where the first (second) column denotes the BC's at $y=0\   (y=L) $.
In the first two models of type (N,N) and (D,D), there appears 
a massless mode and other KK modes acquire masses on the order of $\frac{1}{L}$.
Thus, those models have nothing interesting.
In the third model of type (D,N), there are no massless modes 
and all KK modes acquire masses on the order of $\frac{1}{L}$.
This model is not a desired one.
The last model of type (N,D) has no massless modes, 
but the lowest mode acquires a tiny mass 
\begin{equation}
	m \sim \frac{2\kappa}{L} e^{-\kappa}.
\end{equation}
The remaining KK modes acquire masses on the order of $\frac{1}{L}$.
Thus, the hierarchical structure in the spectrum is realized in this model.
Since there is no massless mode, the 4d gauge invariance is broken, 
and the mass of the lightest gauge boson is given by (9). 
It is interesting to note that the supercharge $Q$ is still well defined 
but there is no zero energy state with $Q| 0 \rangle=0$ in this model.
Therefore, we may say that supersymmetry is \lq \lq spontaneously" broken.
This observation may allow us to say that gauge symmetry is 
\lq \lq spontaneously" broken in our mechanism, 
though the gauge symmetry breaking is caused by boundary effects.

In summary, we proposed a mechanism to break gauge symmetry 
with mass scales much smaller than any scales of underlying 
higher dimensional gauge theories without a fine tuning.
A key observation in our formulation is that the problem of 
gauge symmetry breaking can be translated into that of supersymmetry breaking.
It would be interesting to construct realistic GUT models 
without the hierarchy problem along the line discussed here.


\baselineskip 5mm 

%
%
%
\end{document}